\documentclass[11pt]{article}

\usepackage{graphicx}
\usepackage{amsmath}
\usepackage{amssymb}
\usepackage{bm}
\usepackage{mathtools}
\usepackage{stfloats}

\begin{document}


\title{Is cortical criticality unique?}

\author{Karlis Kanders and Ruedi Stoop, \\
Institute of Neuroinformatics and Institute of Computational Science,\\
University of Zurich and ETH Zurich, \\ Winterthurerstr. 190, \\ 8057 Zurich, Switzerland\\
}

\date{\today}



\maketitle

\begin{abstract}
There are indications that for optimizing neural computation, neural networks - including the brain -   operate at criticality. Previous approaches have, however, used diverse fingerprints of criticality, leaving open the question whether they refer to a unique critical point or whether there could be several.  
Using a recurrent spiking neural network as the model, we demonstrate that avalanche criticality does not necessarily lie at the dynamical edge-of-chaos and that therefore, the different fingerprints indicate distinct phenomena with an as yet unclarified relationship.
\end{abstract}


\begin{figure}[h!!!]
\includegraphics[width=1.0\textwidth]{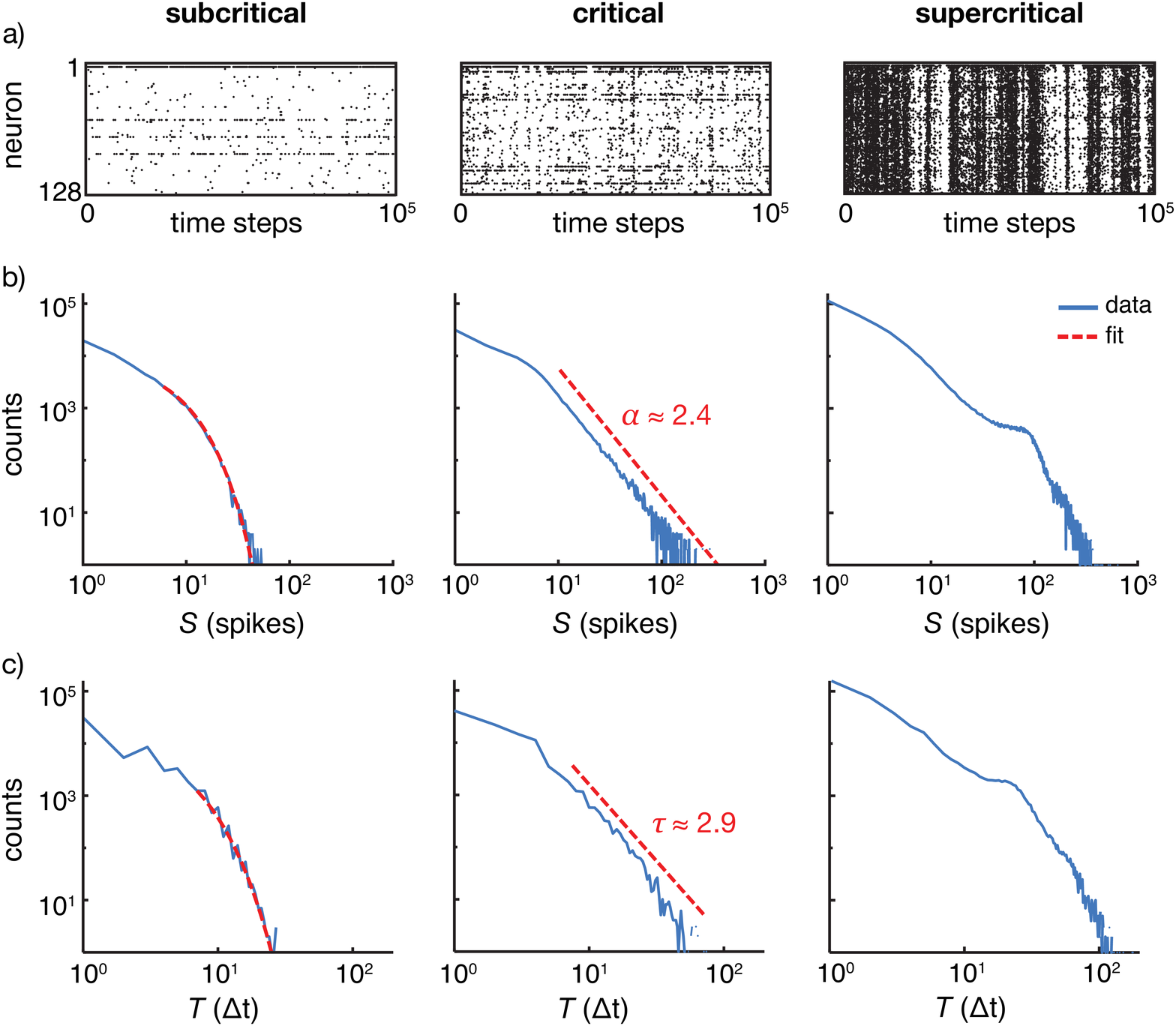}
\caption{Evolution of the topological network state from subcritical (left panel), to critical (middle), to supercritical (right) upon increasing the synaptic weight scaling parameter $W$ from 0.13, to 0.139 and to 0.15, respectively. (a) Raster plots of spiking activity (different weight configurations underlie each plot). First row: activity of the intrinsically spiking neuron. (b) Avalanche size $S$ distributions. (c) Avalanche lifetime $T$ distributions. Red dashed lines show maximum likelihood fits. The critical network exhibits for $S \in [6, 100]$, a power law size distribution with an exponent $\alpha = 2.41$ (p-value = 0.52). The lifetime distribution across the same interval yields an exponent $\alpha = 2.93$ (p-value = 0.38). The subcritical size and lifetime distributions follow an exponential decay with the decay constant $\mu = 0.21$ (p-value = 0.26) and $\mu = 0.40$ (p-value = 0.09), respectively.}
\label{distributions}
\end{figure}

\section{Introduction}

{\it Scale-free avalanches of activity of neuronal firing of biological neural network experiments, have suggested that those networks might be operating at (`topological') phase transitions. Theoretical studies of the dynamics exhibited by artificial neural networks and cellular automata have highlighted potential computational benefits of edge-of-chaos dynamics, i.e., when a system is at a transition between stable and chaotic dynamics, and, in particular, if this transition is of second order. Here, we scrutinize  whether these two phenomena coincide for recurrent neural networks when the spiking neurons are more realistic than in  previous approaches. Using such neurons, we tune the network to subcritical, critical and supercritical topological states, and calculate the Lyapunov exponents of the network. We find that in all three cases, the network exhibits a positive largest Lyapunov exponent indicating chaotic dynamics. This indicates that avalanche criticality does not need to coincide with edge-of-chaos.}\\

In the endeavour of understanding the functioning of the brain, the hypothesis has emerged that biological neural networks might be operating at criticality \cite{Beggs2003, Chialvo2010, Bialek2011}. The promise of this hypothesis is that at the critical point the particular details of the system's individual elements and their interaction laws cease to be of importance \cite{Stanley1987}. In this case, the phase transition itself dominates the behavior of the system and therefore the astounding anatomical and biophysical details of neural circuits would surrender to some very generic network properties, allowing to grasp the fundamentals of the information processing and computation. In addition, several computational advantages of criticality that render such a state particularly attractive have been exhibited, such as optimised information transmission and capacity, increased flexibility of responses granted by diverse activity patterns \cite{Haldeman2005,Shew2011}, and more.
A ``fingerprint" of criticality is power law distributions of the properties exhibited by local descriptors when evaluated across the ensemble. Such  fingerprints were discovered in the statistics of spontaneous activity avalanches of cortical neural tissue recorded with multi-electrode arrays \cite{Beggs2003, Mazzoni2007, Tetzlaff2010} and, more recently, in the auditory system \cite{Stoop2016}. 

However, also competing explanations for power-law like distributions have been provided \cite{Berger2015,Touboul2016}, and, occasionally, even their  emergence in the experimental context has been questioned \cite{Dehghani2012}. As a result, the  \textit{avalanche criticality} hypothesis  \cite{Beggs2003,Levina2007} is still controversial. 
One major reason behind this controversy may be that the notion of computation used in this avalanche or `topological' characterization usually remains vaguely defined. This is similar to the situation on the `dynamical' counterpart, where it has been claimed that `computation' (in the sense of the ability `to transmit, store and modify information' \cite{Langton1990}), would be optimal at \textit{edge-of-chaos} \cite{Langton1990} criticality. 
In recurrent neural networks, edge-of-chaos has most recently been studied in the reservoir computing framework. For best task performance (in the above sense of `computation'), a network requires properties somewhat analogous to the ones ascribed to avalanche criticality: the flexibility to represent spatiotemporally diverse inputs while preserving their distance relationships (i.e., similar inputs should give rise to similar responses). This is indeed provided (and should not come as a surprise) at the system's transition from stable to chaotic dynamics, which is characterised by a vanishing largest Lyapunov exponent \cite{Bertschinger2004, Legenstein2007}. Unfortunately, in the sense of computation as reduction of complexity \cite{Stoop2004}, such a ``reservoir" is not actually computing; it rather serves as a high-dimensional representation space of spatiotemporal input patterns from which the readout neurons can sample, and do the computation. Also, it should be mentioned, that a zero Lyapunov exponent can only be seen as a necessary, but never as a sufficient characterization of the situation underlying  representational (let alone computational) optimality.

Links have occasionally been drawn in the literature between edge-of-chaos and avalanche criticality \cite{Beggs2008, Hesse2014}, however, the precise relationship between the topological and the dynamical criticality is still far from settled. Indeed, the few studies that have demonstrated simultaneous presence of both phase transitions \cite{Haldeman2005, Magnasco2009} have used simplified network models with nodes that had no intrinsic dynamics. This is why in our contribution we examine whether in a recurrent spiking neural network model with realistic, non-trivial node dynamics, avalanche criticality in fact lies at the edge-of-chaos. We will show that this does not need to be the case.

\begin{figure}[h!!]
\includegraphics[width=1.0\textwidth]{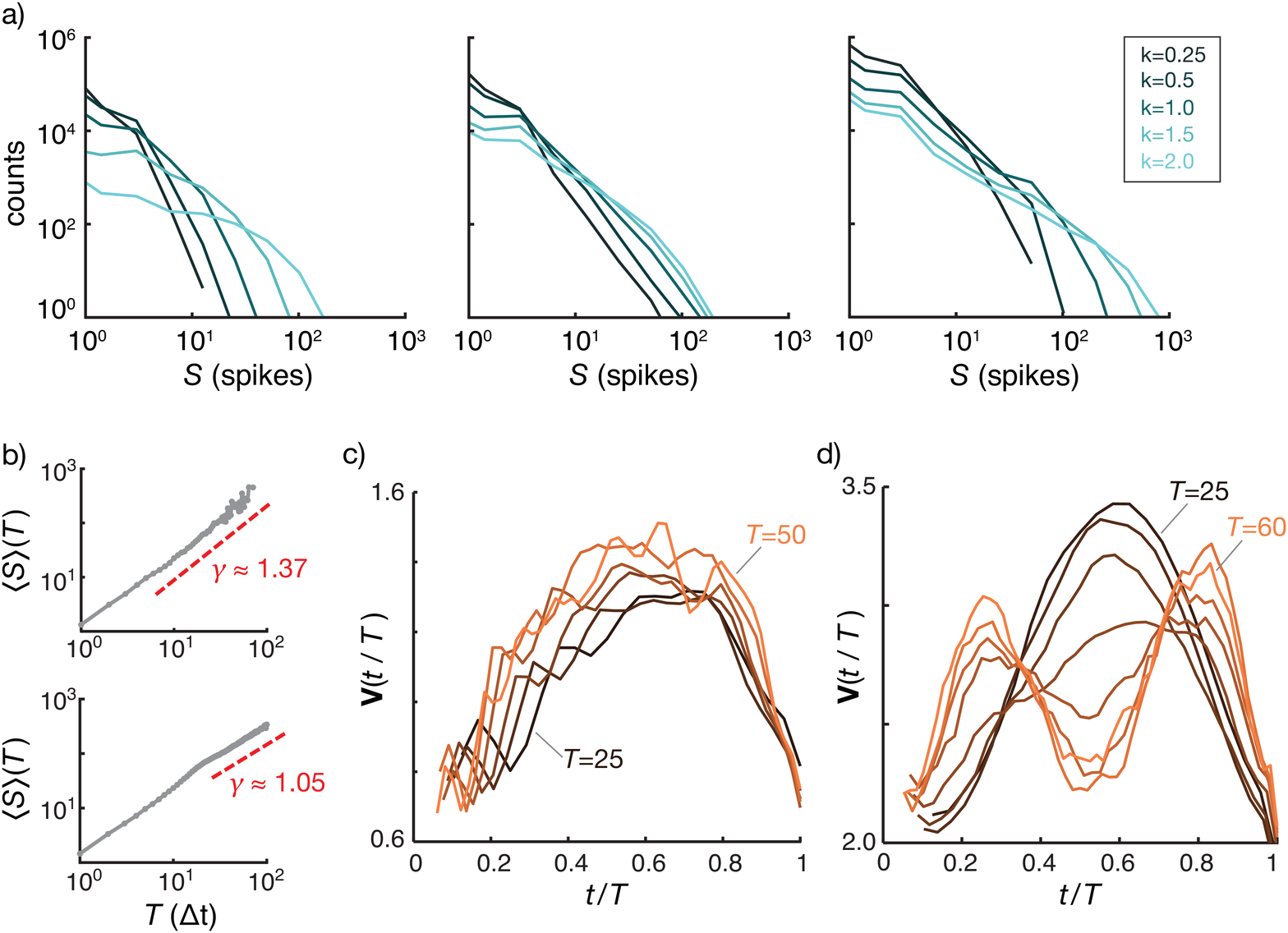}
\caption{Criticality tests. (a) Avalanche size distributions obtained with various temporal bin sizes, $\Delta t=k\cdot \langle IEI \rangle$, for subcritical (left), critical (middle) and supercritical (right) networks. (b) Mean avalanche size as a function of lifetime for critical (top) and supercritical (bottom) networks. The red, dashed line demonstrates a power law relationship $\langle S \rangle(T) \propto T^{\gamma}$. (c)~Critical network's avalanche shapes exhibited a noisy collapse, manifesting a high degree of self-similarity ($T$ = 25,30,$\ldots$,50 from darker to lighter color). (d) Rescaled avalanche shapes of the supercritical network did not exhibit a collapse ($T$ = 25,30,$\ldots$,60).}
\label{tests}
\end{figure}

\section{Neural network model}

Our recurrent neural network model is based on Rulkov model neurons \cite{Rulkov2002}, arranged on an Erd\"{o}s-R\'{e}nyi directed random graph (see Appendix for details). Rulkov's phenomenological, map-based neuron model can reproduce practically all experimentally observed neuronal spiking patterns \cite{Rulkov2004}, and even finer neurobiological details, namely, the phase response curves of real neurons \cite{Kanders2015}. By this feature, our network model differs substantially from the previous efforts of linking avalanche criticality and edge-of-chaos by probabilistic binary units \cite{Haldeman2005} and analog rate neurons \cite{Magnasco2009}. Our network model reflects the general features ascribed to cortical networks: it consists of both excitatory (80\%) and inhibitory (20\%) neurons, its connectivity is sparse (4\%) and inhibitory synapses are several times stronger than the excitatory ones. The network's random topology is more in line with \textit{in vitro} dissociated neural cultures \cite{Eckmann2007} that  so far have provided the strongest evidence for critical avalanches of spiking activity. Spontaneous activity is modelled by setting one of the neurons to spike intrinsically, and provide input to the rest of the network. Hence, the main source of spontaneous activity is embedded into the network, and can therefore be influenced by the network state. In addition, every neuron receives a sparse external excitatory input in the form of independent Poisson spike trains, which models randomness similar to the spontaneous neurotransmitter vesicle release. The parameters of the network model are kept fixed, except for a synaptic weight scaling parameter $W$, that is varied for accessing subcritical, critical and supercritical network activity states.

\section{Avalanche criticality}

To put our study in the same context as the experimental investigations, we follow the established approach of defining neural avalanches as periods of uninterrupted neural activity with respect to a time binning $\Delta t = \langle IEI \rangle$, where $\langle IEI \rangle$ is the average inter-event interval, i.e., the average time between two subsequent spikes in the network \cite{Beggs2003}. The avalanche's size, $S$, is measured as the total number of spikes within the avalanche; the avalanche lifetime, $T$, is the number of time bins that the avalanche spans. The parameters characterizing avalanche size and lifetime  distributions, respectively, are estimated using maximum likelihood, and the goodness of fit is evaluated with the Kolmogorov-Smirnov test \cite{Deluca2013} (see Appendix). 

Upon an increase of the global synaptic connection strength, we observe an overall increase in network activity and an evolution of the topological network state from subcritical, to critical, to supercritical (Fig.~\ref{distributions}). The average inter-event intervals, $\langle IEI \rangle$, are 110, 48, and 8 time steps, for the subcritical, critical and supercritical networks, respectively. The avalanche size distribution of the critical network follows a power law, $P(S) \propto S^{-\alpha}$, with exponent $\alpha \approx 2.4$ (Fig.~\ref{distributions}(b)). The noise cut-off at around $S \approx 100$ must be expected, because the network's  finite size of $N = 128$ elements; in most of the avalanches a single neuron fires only once. In the subcritical case, the avalanches are smaller and their size decays exponentially, while in the supercritical case there is an increased number of large avalanches, signaled by the hump at the end of the distribution. If the synaptic strengths are increased further, the hump becomes even more prominent. A similar metamorphosis of the distribution shape is observed for avalanche lifetimes. At criticality, the lifetime distribution can be fitted with a power law (exponent $\tau \approx 2.9$, Fig.~\ref{distributions}(c)). That fit is somewhat less convincing than the one for the size distribution, which is, however, a commonly observed phenomenon in  electrophysiological experiments \cite{Beggs2003,Pasquale2008, Hahn2010}.

Power law distributions can be caused by several different mechanisms and do as their origin not necessarily request a phase transition. To confirm that the network is truly at criticality  necessitates additional tests: If truly scale-free, the choice of the temporal bin size, $\Delta t$, should not affect the avalanche size distribution (cf. models of self-organized criticality \cite{Priesemann2014}). Fig.~\ref{tests}(a) shows that the size distribution of the critical network is only mildly influenced by different choices of $\Delta t$ and that the effect is markedly smaller compared to the subcritical and supercritical cases.

We further assessed whether, after time-rescaling, the avalanches of the critical network collapse to one characteristic shape  \cite{Sethna2001, Friedman2012}. To this end, the shape of an avalanche with a lifetime $T$ is defined as the temporal evolution of its size, $V(T,t)$. The duration of the avalanches is normalised to $t/T$, and the average avalanche shape for each $T$ is calculated and rescaled to $\textbf{V}(t/T) = T^{1-\gamma} \langle V \rangle(T,t/T)$, where $\langle V \rangle(T,t/T)$ is the average avalanche shape and $\textbf{V}(t/T)$ is the universal scaling function, i.e., the characteristic shape of all avalanches. The critical exponent $\gamma$ is then obtained from the relationship
$\langle S \rangle(T) \propto T^{\gamma}$,
where $\langle S \rangle(T)$ is the mean size of avalanches as a function of their lifetime $T$. Fig.~\ref{tests}(b) shows that the critical network's $\langle S \rangle(T)$ follows a power law relationship, with the exponent $\gamma \approx 1.37$. In the case of the supercritical network, a certain, smaller range of the function also follows a power law, which permits the comparison between the self-similarities of the avalanche shapes of the critical and supercritical states. At the critical point we observe a noisy collapse of the avalanche shapes of duration $T \geq 25$ (Fig. \ref{tests}(c)), and the shapes are considerably more self-similar than in the case of the supercritical network (Fig. \ref{tests}(d)). The avalanche shapes of short lifetimes ($T < 25$) generally fail to collapse. Note that universal scaling at shortest length scales cannot be generically expected, because there the particular behavior of individual system parts can be stronger than the collective behavior \cite{Sethna2001}. 

As a final test we examined whether the crackling noise relationship \cite{Sethna2001, Friedman2012} between critical exponents, $(\tau~-~1)/(\alpha~-~1)~=~\gamma$, holds for our critical network. The critical exponents of avalanche lifetime distribution ($\tau = 2.93$), avalanche size distribution ($\alpha = 2.41$), and the function of the mean avalanche size depending on the lifetime ($\gamma = 1.37$) fulfil this relation. Taken together: power law distributions, the self-similarity of avalanche shapes, and an excellent fulfillment of the fundamental relation between critical exponents, strongly suggest that our `critically tuned' network is indeed critical.

\section{Lyapunov spectra}

\begin{figure}[h!!!!]
\includegraphics[width=1.0\textwidth]{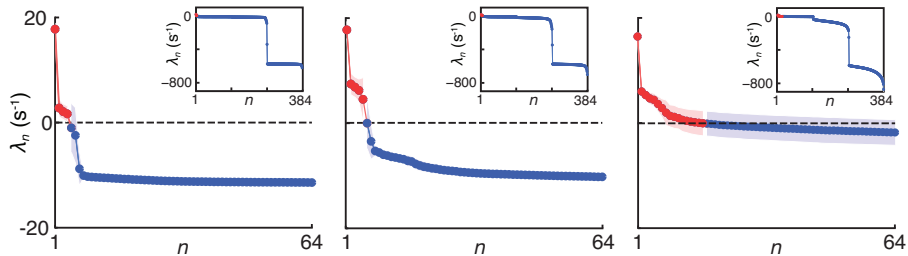}
\caption{Lyapunov spectra of subcritical (left), critical (middle) and supercritical (right) networks, showing the first 64 Lyapunov exponents. Positive values of $\lambda_n$ are shown by red, negative ones by blue circles. Areas within one standard deviation are shaded. Insets: full Lyapunov spectra.}
\label{lyapunov}
\end{figure}

To determine whether avalanche criticality resides at the edge-of-chaos, we calculated the Lyapunov spectrum for the subcritical, critical and supercritical cases by using the Jacobian matrix evaluated at points along the trajectory of the network's state vector \cite{Stoop1988} (see Appendix for details). For avalanche criticality coinciding with edge-of-chaos criticality, we would expect the largest Lyapunov exponent $\lambda_1$, to be negative for the subcritical network, positive for the supercritical network, and vanishing for the critical network. However, $\lambda_1$ turns out to be positive in all of the three cases (Fig.~\ref{lyapunov}). For the subcritical and critical networks it is practically of the same size ($\lambda_1$ = 17.8 s$^{-1}$), whereas it is slightly smaller for the supercritical network ($\lambda_1$ = 16.4 s$^{-1}$, where the units of Lyapunov exponents were obtained by applying time rescaling \cite{Rulkov2004} with one time step accounting for 0.5 ms). Lyapunov spectra provide some help to understand this phenomenon: Upon increased synaptic strength, the total number of positive Lyapunov exponents increases. Every positive Lyapunov exponent amplifies perturbations of the microstate to an observable change in the macrostate. The sum of all positive $\lambda_n$ gives the total average rate of this amplification, $H=\sum_{\lambda_n>0} \lambda_n$. This sum is also known as the upper bound of the Kolmogorov-Sinai entropy \cite{Peinke1992}, and can be interpreted as the entropy production rate. $H$ increases with stronger synaptic coupling: from 28$\pm$6 s$^{-1}$ (mean$\pm$standard deviation) for the subcritical case, to 46$\pm$12 s$^{-1}$ for the critical case, to 88$\pm$54 s$^{-1}$ for the supercritical case. Therefore, although the supercritical network has a slightly smaller largest Lyapunov exponent, it loses information about a past state at a faster rate.\\
Chaotic dynamics could be a collective effect of the network interactions, or arise simply because nodes themselves have chaotic dynamics. To check this, we measured the largest Lyapunov exponent of the intrinsically spiking neuron in the absence of network input, and found it to be positive ($\lambda_1$ = 20 s$^{-1}$). In the presence of external input, the neuron was occasionally silenced (see also Fig.~\ref{distributions}(a)), which is a behavior that is often observed in Class II neurons \cite{Prescott2008}  in the vicinity of a Andronov-Hopf bifurcation from the resting state to spiking \cite{Guttman1980}. The intrinsically spiking Rulkov's neuron used in our simulations is close to the Andronov-Hopf bifurcation, and some perturbations are able to push the neuron's state variable close to the unstable fixed point. During the time it takes to escape from the fixed point, the neuron does not fire. As a result of this occasional silencing, the neuron's largest Lyapunov exponent drops to $\lambda_1$ = 18 s$^{-1}$, which is in agreement with the value of $\lambda_1$ found in our network analysis, suggesting that the largest Lyapunov exponent of the network might be capturing the dynamics of the intrinsically spiking neuron. In the subcritical and critical cases there are, however, 3-4 more positive Lyapunov exponents, which is close to the number of neurons that receive inputs from the intrinsically spiking neuron. Therefore, the source of chaos in our networks resides in the single neuron dynamics; the increase of coupling strength made the chaos more intensive because more neurons are pushed towards spiking.

\section{Discussion}

Our study shows that avalanche criticality does not need to co-exist with edge-to-chaos criticality. For all choices of the synaptic weights, the network exhibits chaotic dynamics, even though qualitative changes in the topological network state are observable. This result suggests that in neural networks with non-trivial node dynamics we may have two separate phase transitions. 
As a consequence of this finding, we suggest that for the investigations of the computational effects of the topological network state, the network's dynamical state should also be taken into account, if not given priority. Results regarding avalanche criticality obtained in a dynamically stable network, might no longer be relevant for a chaotic network with its unpredictable activity patterns. In addition, the present study also provides an interesting paradox that may bear importance for understanding biological network behavior: Upon an increase of the synaptic coupling, chaos may intensify in the sense of a larger entropy production rate, whereas the spiking can become more correlated in time, resulting in an increase of both ``order" and chaos.

The critical exponent of our avalanche size distribution, $\alpha = 2.4$, differs from $\alpha = 1.5$ measured in the original analysis of local field potential avalanches in Ref. \cite{Beggs2003}. However, subsequent works on critical spike avalanches have seen substantially varied exponents, covering a range from 1.5 to 2.1 \cite{Mazzoni2007, Tetzlaff2010}. Our size distribution at criticality is similar to the distributions reported for dissociated rat cortical cultures in Ref. \cite{Tetzlaff2010} with $\alpha = 2.1$ at the tail of the distribution, and another scaling regime for very small avalanches. An exponent of $\alpha = 2.5$ has been found for ``background" avalanches in simulations of bursting recurrent networks \cite{Orlandi2013}. This exponent has been linked to critical percolation on a Cayley tree-like network, that also yields the same exponent for the cluster size distribution \cite{Stauffer1994}. This also suggests that there is more than one avalanche criticality state in recurrent neural networks. Because of the insufficient experimental and theoretical considerations \cite{Priesemann2014}, some researchers have explicitly rejected the idea of a universal exponent for neural avalanches. Avalanches with exponent $\alpha = 1.5$ originally obtained for critical branching processes, were recently also measured for cochlear activation networks \cite{Stoop2016} and might, therefore, have a more general significance in the context of neural computation.

Our investigations were based on a well-defined, precise calculation method of the network's Lyapunov exponents. The majority of the studies on dynamical stability in neural networks have used the perturbation method only, i.e., by repeating every simulation of the network activity after adding a random perturbation $d_0$ to the state vector, so that the largest Lyapunov exponent could be assessed from the evolution of the distance between the network's unperturbed and perturbed trajectories \cite{Haldeman2005, Bertschinger2004, Legenstein2007, Zhou2009}. This approach, however, only gives an estimate of $\lambda_1$ and does not provide information about the rest of the Lyapunov exponents. Some studies have employed an approach of using ``spike-sized" perturbations \cite{Bertschinger2004, Legenstein2007}. This, however, may be too far away from the perturbation limit of $d_0 \to 0$ that is relevant for the definition of Lyapunov exponents. It could, therefore, theoretically, not be excluded that the largest Lyapunov exponent obtained following such approaches, depends on the size of the perturbation, so that instead of the positive values of the first Lyapunov exponent, in the limit of infinitely small perturbations a negative value might emerge \cite{Monteforte2012}. The method employed here will not suffer from such potential shortcomings and has been shown to provide richer insight into the network dynamics.

\section{Acknowledgements} 
We would like to express our gratitude to Tom Lorimer for critical discussion and reading of the manuscript.
This work was supported by the Swiss National Science Foundation grant  200021 153542/1 and by an internal grant of ETHZ, ETH-37 152.



\appendix*
\section{METHOD DETAILS}

\subsection{Neuron model and network parameters}

The number of neurons in the network was set to $N$ = 128, and every neuron was assigned random $k = 5$ postsynaptic neighbours, which resulted in approximately $4\%$ connectivity. The size of the network was chosen to satisfy a trade-off between obtaining enough statistics for the avalanche size distributions and minimizing the calculation time of the network's Lyapunov exponents. 80\% of the neurons were excitatory and 20\% inhibitory.

We modelled neuron dynamics using Rulkov's two-dimensional map: \cite{Rulkov2002, Rulkov2004}
\begin{subequations}
\begin{align}
    x_{n+1} &= 
    \begin{dcases}
    \frac{\alpha}{1-x_{n}} + u & x_n\leq 0,\\
    \alpha + u & 0 < x_n < \alpha + u \wedge x_{n-1}\leq 0,\\
    -1 & x_n \geq \alpha + u \lor x_{n-1} > 0,
    \end{dcases}\\ \nonumber
y_{n+1} &= y_{n} - \mu(1+x_{n}) + \mu\sigma + \mu I^{syn}_n,\\
\end{align}
\end{subequations}
where $u =  y_{n} + \beta I^{syn}_n$. From resting to spiking state, this map undergoes a Hopf bifurcation, which corresponds to a Class II neuron type \cite{Prescott2008}. The parameter values for excitatory and inhibitory neurons were identical: $\alpha = 3.6$, $\mu = 0.001$, $\sigma=0.09$, $\beta = 0.133$. With $\sigma = 0.103$, the intrinsically spiking neuron was poised just above the spiking threshold.

Synaptic input $I^{syn}_{n}$ was modelled by an exponential decay and step-like increase upon a presynaptic spike event (denoted as spike$_j$):
\begin{equation}
    I^{syn}_{n+1} = \eta I^{syn}_{n} -
    \begin{dcases}
    \sum^N_j W w_{ij}(x_{n} - x_{rp}) & \text{spike}_j,\\
    0 & \text{otherwise,}
    \end{dcases}
\label{eq_synapse}
\end{equation}
where $\eta$ controls the decay rate of the synaptic current, $w_{ij}$ is the synaptic strength between the presynaptic neuron $j$ and the postsynaptic neuron $i$, $x_{rp}$ is the reversal potential which determines whether the synapse is inhibitory or excitatory, and $W$ is a global scaling parameter of the synaptic weight. We used the following parameter values for excitatory (`Ex') and inhibitory (`Inh') synapses: $x_{rp}^{Ex} = 0$, $\eta^{Ex} = 0.75$, $w_{ij}^{Ex} = 0.6$, $x_{rp}^{Inh} = -1.1$, $\eta^{Inh} = 0.75$, $w_{ij}^{Inh} = 1.8$. The probability for a single neuron to receive an external input spike at any given iteration was $6 \cdot 10^{-4}$. Synaptic plasticity was not included in the network model, since self-organized criticality is not an issue of the present investigation. Instead, the synaptic weight scaling parameter $W$ was varied, for finding subcritical, critical and supercritical activity states. For each of the three states we ran 50 simulations and pooled the results. For each simulation, the synaptic connections were randomized. A single simulation covered $5 \cdot 10^5$ time steps, where the first 5000 steps were discarded.

\subsection{Estimation of the distribution parameters}

The theoretical fits to the avalanche size and lifetime distributions were found using the guidelines in Ref. \cite{Deluca2013}. We assumed that the observations $\mathbf{s}$ (avalanche size or lifetime) were sampled independently from a distribution $p(s|\alpha)$ parametrised by $\alpha$. The likelihood of the parameter $\alpha$ is given by the probability of the observations $\mathbf{s}$, given $\alpha$:
\begin{equation}
L(\alpha|\mathbf{s}) = \prod_{i=1}^N p(s_i|\alpha),
\end{equation}
where N is the number of samples. In practice, we used the logarithm of the likelihood, $\ell(\alpha|\mathbf{s}) = \text{ln} L(\alpha|\mathbf{s})$, which allows to replace the product with a sum. Log-likelihood has a maximum at the same $\alpha$ as the likelihood, due to monotonic nature of the logarithm function. Thus the maximum likelihood estimator of the parameter $\alpha$ is
\begin{equation}
\hat{\alpha} = \text{argmax}_{\alpha} \ \ell(\alpha|\mathbf{s})  = \text{argmax}_{\alpha} \ \sum_{i=1}^N \text{ln} \ p(s_i|\alpha).
\end{equation}
In the case of a discrete, truncated power law distribution of $s$ with the scaling exponent $\alpha$, within the bounds $s_{min} = a$ and $s_{max} = b$, the probability of $s_i$ is
\begin{equation}
p(s_i|\alpha) = \frac{s_i^{-\alpha}}{\sum_{j=a}^b j^{-\alpha}},
\end{equation}
which gives the log-likelihood that needs to be maximised:
\begin{equation}
\ell(\alpha|\mathbf{s}) = -\alpha \sum_{i=1}^N \text{ln} s_i - N \ \text{ln} \ \sum_{j=a}^b j^{-\alpha}.
\end{equation}
Similarly we can derive the log-likelihood of the exponential decay constant, $\mu$, of a discrete, truncated exponential distribution of $s$:
\begin{equation}
\ell(\mu|\mathbf{s}) = - \mu \frac{1}{N} \cdot \sum_{i=a}^b s_i -\text{ln} \ \sum_{j=a}^b e^{-\mu j}.
\end{equation}

Goodness of fit was evaluated using the p-value of Kolmogorov-Smirnov (KS) distance. KS distance, $d_{KS}$ is the maximum difference in absolute value between the empirical (avalanche size or lifetime) survival function $S_e(s)$ and the theoretical one $S(s|\hat{\alpha})$. For a good fit, $d_{KS}$ should be small, but the relative scale of $d_{KS}$ is given by its own probability distribution. The p-value provides the probability that the KS distance takes a value larger than the one obtained empirically. We generated 1000 synthetic datasets using the estimated parameter $\hat{\alpha}$ and the same number of samples as in the empirical distribution. The parameters of the synthetic distributions were estimated using maximum likelihood and $d_{KS}$ was calculated for each synthetic fit. The p-value is given by the fraction of synthetic fits which have a higher $d_{KS}$ (worse fit) than the empirical fit. A threshold of p-value $>$ 0.05 was chosen as the criterion for accepting a fit. 

\subsection{Avalanche shapes}

\begin{figure}[h!!!!!]
\includegraphics[width=\linewidth]{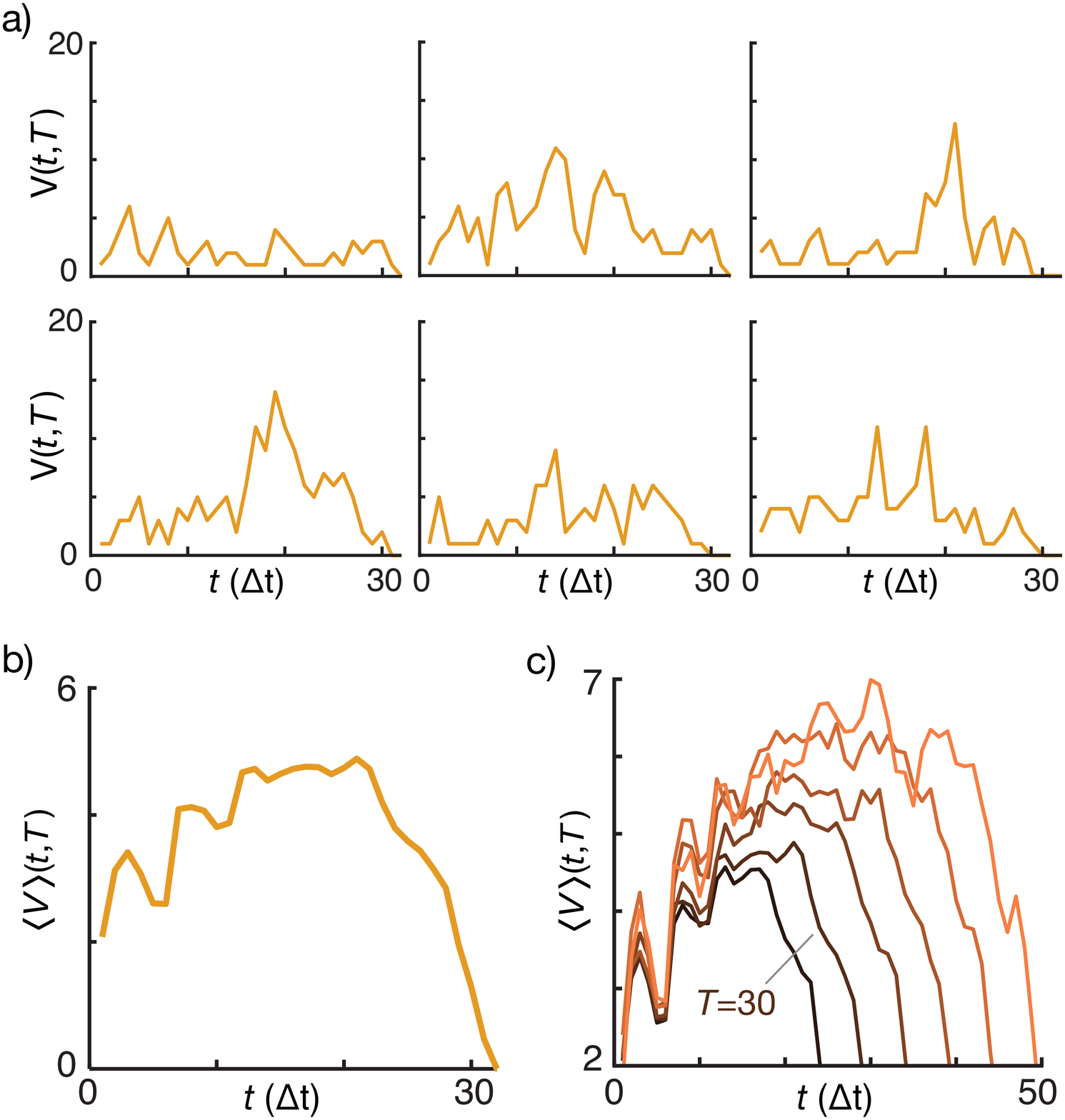}
\caption{Avalanche shapes of the critical network. (a) Individual avalanches had a highly variable shape: examples of the temporal profiles (spikes vs time) of avalanches with a duration T = 30$\pm$2. (b) Mean shape for an avalanche of duration T = 30, calculated from 650 samples. (c) Average avalanche shapes of durations T = 25,30,$\ldots$,50.}
\label{shapes}
\end{figure}

Individual avalanche shapes, $V(t, T)$, were highly variable and the average over many avalanche samples was calculated to get an estimate of the mean avalanche shape, $\langle V \rangle (t, T)$ (Fig.~\ref{shapes}). To increase the sample size and obtain better defined averages, we included all avalanches with a lifetime $T\pm2$ in the calculation of $\langle V \rangle (t, T)$. For the same reason, we also performed 150 additional simulations for the critical network. The number of avalanche shape samples for each lifetime ranged between 100--7500, with larger avalanches having a smaller number of samples.

\subsection{Calculation of the Lyapunov spectrum}

The largest Lyapunov exponent $\lambda_1$ describing the average rate of the exponential separation of the system's trajectories, is used to determine whether a dynamical system is stable or chaotic. Accordingly, the distance between trajectories evolves exponentially over time as $d_n = d_0 e^{\lambda_1 n}$, in the limit $d_0 \to 0$. For $\lambda_1 < 0$, nearby trajectories converge on average exponentially, while $\lambda_1 > 0$ implies divergence of nearby trajectories and hall-marks chaos. At the critical point, $\lambda_1 = 0$; in its neighborhood, the system experiences a critical slowing down of the dynamics, where small perturbations can have long-lasting effects. We numerically determined $\lambda_1$ using the local linearisation along the system's trajectory, i.e., the Jacobian matrix of the neural network \cite{Stoop1988, Peinke1992}. This powerful method not only yields $\lambda_1$, but provides the whole Lyapunov spectrum, i.e., all Lyapunov exponents of the system.

Every neuron lives in a three-dimensional state space: the two state variables $x_n$ and $y_n$, and the synaptic input variable $I^{syn}_{n}$. The Jacobian matrix for a single neuron has the form

\begin{equation}
\centering
    J_n =
    \begin{dcases}
        \begin{bmatrix}
            \frac{\alpha}{(1-x_n)^2} &1  &\beta \\ 
            -\mu &1  &\mu \\ 
            -\sum^N_j w_{ij}\text{*} &0  &\eta 
        \end{bmatrix} & x_n\leq 0,\\
        \\
        \begin{bmatrix}
            0 &1  &\beta \\ 
            -\mu &1  &\mu \\ 
            -\sum^N_j w_{ij}\text{*} &0  &\eta 
        \end{bmatrix} & 0 < x_n < \alpha + u \wedge x_{n-1}\leq 0,\\
        \\
        \begin{bmatrix}
            0 &0  &0 \\ 
            -\mu &1  &\mu \\ 
            -\sum^N_j w_{ij}\text{*} &0  &\eta 
        \end{bmatrix} & x_n \geq \alpha + u \lor x_{n-1} > 0,\\
    \end{dcases}
\end{equation}
where $u =  y_{n} + \beta I^{syn}_n$ and * denotes the case of presynaptic spike events from neurons $j$ (otherwise, the corresponding entry vanishes). The extension of the single neuron Jacobian matrix to the full network is straight-forward: the state variables of a neuron do not directly depend on the state variables of other neurons because the interaction is only through spike events and we can write the Jacobian of the full network, $J^{net}_n$, as a $3N \times 3N$ block diagonal matrix with the Jacobians of the individual neurons on the diagonal and all other elements being equal to 0.

The Lyapunov exponents were obtained by following the ellipsoid $J^{net}_n O$ (where $O$ is the unit sphere) as it grows one iteration at a time. After every time step Gram-Schmidt orthonormalisation procedure was applied to acquire a set of orthogonal vectors $\{\mathbf{y}_1,...,\mathbf{y}_{3N}\}$. The orthonormalisation procedure is computationally expensive, which makes the calculation of Lyapunov exponents for large networks slow. One step growth in direction $i$ is captured by the length of the $i\text{-}th$ orthogonal vector $\left \| \mathbf{y}_i \right \|$. The total expansion can be written as $r_i^n =   \left \| \mathbf{y}_i^1 \right \| \left \| \mathbf{y}_i^2 \right \| \hdots \left \| \mathbf{y}_i^n \right \|$. We assume that the expansion grows exponentially $r_i^n = e^{\lambda_i n}$ and therefore the rate of expansion in the $i\text{-}th$ direction, i.e., the Lyapunov exponent $\lambda_i$, can be calculated as
\begin{equation}
\lambda_i = \frac{1}{n} \sum^n_{t=1}\text{ln}\left \| y_i^t \right \|.
\end{equation}
Note that in the main text we use  for the direction number subscript $n$, instead of $i$.

We calculated the Lyapunov exponents of the subcritical, critical and supercritical networks for 10 random configurations out of the 50 that we used to obtain the avalanche size and lifetime distributions. The simulation length for the calculations was kept at $7.5 \cdot 10^4$ time steps. The Lyapunov exponents converged well, but, to take care of potential fluctuations, the final value of $\lambda_n$ was obtained by averaging over the last 5000 steps.



\begin{thebibliography}{9}

\bibitem{Beggs2003} J.~M. Beggs and D. Plenz, ``Neuronal avalanches in neocortical circuits," J. Neurosci. \textbf{23}, 11167-11177 (2003).

\bibitem{Chialvo2010} D.~R. Chialvo. ``Emergent complex neural dynamics," Nat. Phys. \textbf{6}, 744-750 (2010).

\bibitem{Bialek2011} T.~Mora and W.~Bialek, ``Are biological systems poised at criticality?" J. Stat. Phys. \textbf{144}, 268–302 (2011).

\bibitem{Stanley1987} H.~E.~ Stanley,  {\em Introduction to Phase Transitions and Critical Phenomena} (Oxford University
Press, Oxford, 1987).

\bibitem{Shew2011} W.~L. Shew, H. Yang, S. Yu, R. Roy, and D. Plenz ``Information capacity and transmission are maximized in balanced cortical networks with neuronal avalanches," J. Neurosci. \textbf{31}, 55-63 (2011).

\bibitem{Haldeman2005} C. Haldeman and J.~M. Beggs, ``Critical branching captures activity in living neural networks and maximizes the number of metastable states," Phys. Rev. Lett. \textbf{94}, 058101 (2005).


\bibitem{Mazzoni2007} A. Mazzoni, F.~D. Broccard, E. Garcia-Perez, P. Bonifazi, M.~E. Ruaro, and V. Torre, ``On the dynamics of the spontaneous activity in neuronal networks," PLoS ONE \textbf{2}(5), e439 (2007).

\bibitem{Tetzlaff2010} C. Tetzlaff, S. Okujeni, U. Egert, F. W\"{o}rg\"{o}tter, and M. Butz, ``Self-organized criticality in developing neuronal networks," PLoS Comput. Bio. \textbf{6}, e1001013 (2010).

\bibitem{Stoop2016} R. Stoop and F. Gomez, ``Auditory power-law activation avalanches exhibit a fundamental computational ground state," Phys. Rev. Lett. \textbf{117}, 038102 (2016).


\bibitem{Touboul2016} J. Touboul and A. Destexhe, ``Power-law statistics and universal scaling in the absence of criticality," arXiv:1503.08033 (2016).

\bibitem{Berger2015} D. Berger, S. Joo, T. Lorimer, Y. Nam, and R. Stoop, ``Power laws in neuronal culture activity from limited availability of a shared resource," in {\em Emergent Complexity from Nonlinearity, in Physics, Engineering and the Life Sciences} (Springer Proc. Phys., 2017).

\bibitem{Dehghani2012} N. Dehghani, N.~G. Hatsopoulos, Z.~D. Haga, R.~A. Parker, B. Greger, E. Halgren, S.~S. Cash, and A. Destexhe, ``Avalanche analysis from multielectrode ensemble recordings in cat, monkey, and human cerebral cortex during wakefulness and sleep," Front. Physiol. \textbf{3}, 302 (2012).

\bibitem{Levina2007} A. Levina, J.~M. Herrmann, and T.~Geisel, ``Dynamical synapses causing self-organized criticality in neural networks," Nat. Phys. \textbf{3}, 857-860 (2007).


\bibitem{Langton1990} C.~G. Langton, ``Computation at the edge of chaos: phase transitions and emergent computation," Physica D \textbf{42}, 12-37 (1990).

\bibitem{Bertschinger2004} N. Bertschinger and T. Natschl\"{a}ger, ``Real-time computation at the edge of chaos in recurrent neural networks," Neural Comput. \textbf{1436}, 1413-1436 (2004).
 
\bibitem{Legenstein2007} R. Legenstein and W. Maass, ``Edge of chaos and prediction of computational performance for neural circuit models," Neural Networks \textbf{20}, 323-334 (2007).

\bibitem{Stoop2004} R. Stoop and N. Stoop, ``Natural computation measured as a reduction of complexity," \textbf{14}, 675-679 (2004).

\bibitem{Beggs2008} J.~M. Beggs, ``The criticality hypothesis: how local cortical networks might optimize information processing," Phil. Trans. R. Soc. A \textbf{366}, 329-343 (2007).

\bibitem{Hesse2014} J. Hesse and T. Gross, ``Self-organized criticality as a fundamental property of neural systems," Front. Syst. Neurosci. \textbf{8}, 166 (2014).

\bibitem{Magnasco2009} M. Magnasco, O. Piro, and G. Cecchi, ``Self-tuned critical anti-Hebbian networks," Phys. Rev. Lett. \textbf{102}, 258102 (2009).

\bibitem{Rulkov2002} N. Rulkov, ``Modeling of spiking-bursting neural behavior using two-dimensional map," Phys. Rev. E \textbf{65}, 041922 (2002).

\bibitem{Rulkov2004} N.~F. Rulkov, I. Timofeev, and M. Bazhenov, ``Oscillations in large-scale cortical networks: map-based model," J. Comput. Neurosci. \textbf{17}, 203-223 (2004).

\bibitem{Kanders2015} K. Kanders and R. Stoop ``Phase response properties of Rulkov's model neurons," in {\em Emergent Complexity from Nonlinearity, in Physics, Engineering and the Life Sciences} (Springer Proc. Phys., 2017).

\bibitem{Eckmann2007} J.~P. Eckmann, O. Feinerman, L. Gruendlinger, E. Moses, J. Soriano, and T. Tlusty, ``The physics of living neural networks," Phys. Rep. \textbf{449}, 54-76 (2007).

\bibitem{Deluca2013} A. Deluca and \'{A}. Corral, ``Fitting and goodness-of-fit test of non-truncated and truncated power-law distributions," Acta Geophys. \textbf{61}, 1351-1394 (2013).

\bibitem{Pasquale2008} V. Pasquale, P. Massobrio, L.~L. Bologna, L. L., M. Chiappalone, and S. Martinoia, ``Self-organization and neuronal avalanches in networks of dissociated cortical neurons," Neuroscience \textbf{153}, 1354–1369 (2008).

\bibitem{Hahn2010} G. Hahn, T. Petermann, M.~N. Havenith, S. Yu, W. Singer, D. Plenz, and D. Nikolic, ``Neuronal avalanches in spontaneous activity in vivo," J. Neurophysiol. \textbf{104}, 3312–3322 (2010).

\bibitem{Priesemann2014} V. Priesemann, M. Wibral, M. Valderrama, R. Pr\"{o}pper, M. Le Van Quyen, T. Geisel, et al., ``Spike avalanches in vivo suggest a driven, slightly subcritical brain state," Front. Syst. Neurosci. \textbf{8}, 108 (2014).

\bibitem{Sethna2001} J.~P. Sethna, K.~A. Dahmen, and C.~R. Myers, ``Crackling noise," Nature \textbf{410}, 242-250 (2001).

\bibitem{Friedman2012}  N. Friedman, S. Ito, B.~A.~W. Brinkman, M.~Shimono, R.~E.~L. DeVille, K.~A. Dahmen, J.~M. Beggs, and T.~C. Butler, ``Universal critical dynamics in high resolution neuronal avalanche data," Phys. Rev. Lett \textbf{108}, 208102 (2012).

\bibitem{Stoop1988} R. Stoop and P.~F.~ Meier, ``Evaluation of Lyapunov exponents and scaling functions from time series," J. Opt. Soc. Am. B., \textbf{5}, 1037-1045 (1988).

\bibitem{Prescott2008} S.~A. Prescott, Y. DeKoninck, and T.~J. Sejnowski, ``Biophysical basis for three distinct dynamical
mechanisms of action potential initiation," PLoS Comput. Biol. \textbf{4}, e1000198 (2008).


\bibitem{Guttman1980} R. Guttman, S. Lewis, and J. Rinzel, ``Control of repetitive firing in squid axon membrane as a model for a neurone oscillator," J. Physiol. \textbf{305}, 377-395 (1980).


\bibitem{Orlandi2013} J.~G. Orlandi, J. Soriano, E. Alvarez-Lacalle, S. Teller, and J. Casademunt, ``Noise focusing and the emergence of coherent activity in neuronal cultures," Nat. Phys. \textbf{9}, 582-590 (2013).


\bibitem{Stauffer1994} D. Stauffer and A. Aharony, {\em Introduction To Percolation Theory} (Taylor \& Francis, London, 1994).

\bibitem{Zhou2009} D. Zhou, A.~V. Rangan, Y. Sun, and D. Cai, ``Network-induced chaos in integrate-and-fire neuronal ensembles," Phys. Rev. E \textbf{80}, 031918 (2009).

\bibitem{Monteforte2012} M. Monteforte and F. Wolf, ``Dynamic flux tubes form reservoirs of stability in neuronal circuits," Phys. Rev. X \textbf{2}, 041007 (2012).

\bibitem{Peinke1992} J. Peinke, J. Parisi, O.~E. R\"{o}ssler, and R. Stoop, {\em Encounter with chaos: Self-organized hierarchical complexity in semiconductor experiments} (Springer, Berlin, 1992).

\end{thebibliography}
\end{document}